\documentclass[12pt]{article}

\usepackage{graphicx}

\title{ Hadron wave functions in high-energy scattering, form factors and strong decays: the effects
of the Lorentz contracted form}
\author{  Yu.A.Simonov \\
NRC "Kurchatov Institute" - ITEP \\
Moscow, 117218 Russia}

\newcommand{\beq}{\begin{eqnarray}}
 \newcommand{\eeq}{\end{eqnarray}}
\newcommand{\be}{\begin{equation}}
 \newcommand{\ee}{\end{equation}}

\def\fun#1#2{\lower3.6pt\vbox{\baselineskip0pt\lineskip.9pt
\ialign{$\mathsurround=0pt#1\hfil ##\hfil$\crcr#2\crcr\sim\crcr}}}

\newcommand{{\SD}}{\rm SD}

\newcommand{{\Mc}}{\mathcal{M}}

\newcommand{\vex}{\mbox{\boldmath${\rm x}$}}
\newcommand{\vey}{\mbox{\boldmath${\rm y}$}}
\newcommand{\ver}{\mbox{\boldmath${\rm r}$}}

\newcommand{\veP}{\mbox{\boldmath${\rm P}$}}
\newcommand{\vep}{\mbox{\boldmath${\rm p}$}}
\newcommand{\veq}{\mbox{\boldmath${\rm q}$}}
\newcommand{\veQ}{\mbox{\boldmath${\rm Q}$}}
\newcommand{\vez}{\mbox{\boldmath${\rm z}$}}

\newcommand{\veR}{\mbox{\boldmath${\rm R}$}}

\newcommand{\vek}{\mbox{\boldmath${\rm k}$}}

\newcommand{\vev}{\mbox{\boldmath${\rm v}$}}

\newcommand{\vexi}{\mbox{\boldmath${\rm \xi}$}}
\newcommand{\veta}{\mbox{\boldmath${\rm \eta}$}}

\newcommand{\veJ}{\mbox{\boldmath${\rm J}$}}

\newcommand{\lan}{\langle}
\newcommand{\ran}{\rangle}

\begin{document}
\maketitle
\begin{abstract}
We study the class of processes where the dynamics depends essentially on the properties of the hadron wave functions involved in the reactions. In this case the momentum dependence of the form of wave functions imposed by Lorentz invariance and in particular by the Lorentz contraction  can be tested in the experiment and may strongly influence the resulting cross sections. One example of
such observables is given by the hadron form factors, where it was shown that the large Q behavior is mostly frozen, when the
Lorentz  contraction of the hadron wave functions is taken into account. Another example considered previously is the strong hadron decay with high energy emission. In this paper we study the role of the Lorentz contraction in the high-energy hadron-hadron scattering process at large momentum transfer.
For the $pp$ and $p\bar p$ scattering at large $s$ it is shown that at small $-t << s$ the picture of two exponential slopes in the differential cross section explained previously by the author is stable while the backward scattering cross section is strongly increased by the Lorentz contraction.
\end{abstract}

\section{ Introduction}

In the standard field theory operating with point-like objects all interaction vertices are also point-like and
the size and the form of the wave functions do not appear in the matrix elements and cross sections. The exclusion
is however the form factors,where the properties of wave functions are important and actually define the behavior
of form factors. One can envisage also other processes where wave functions can play an important role, such as
the hadron decays and hadron-hadron cross sections with exchange of another hadrons etc. Actually any hadron is
 a complicated system of basic elements considered elementary (quarks,gluons etc.) comprising its bound state plus
additional states in the corresponding Fock column. Therefore the total amplitude is a matrix element containing
integrals of the wave functions of hadrons participating in the process.
 At this point one may ask about the Lorentz (Poincare) invariance of the resulting matrix elements and -- what is more important for us below -- about the possible Lorentz corrected form of the wave functions participating in these
 matrix element. Here one should have in mind that wave functions appearing in the integral in this matrix element
 may be for hadrons with different velocities,so that the overlap integral contains a product of differently Lorentz deformed wave functions. The basic point here is that one should specify the dynamical scheme used in this analysis. Indeed, as it
 was shown by Dirac \cite{1} the dynamical formalism can exploit  1/the instant form 2/ the point form 3/ the light front form, and we shall be using below the instant form implying that the wave function and interaction in the multicomponent relativistic system refers to the instantaneous picture of all constituents. One may ask how this condition
 can be implemented in the standard relativistic QED and QCD formalism.
 The analysis of the Poincare properties of the Hamiltonian and its wave functions was the topic of intense
 activity \cite{2,3,4,5,6,7,8,8*} both in the 1+1 and 3+1 dimensions, see also \cite{8,9,10} for useful reviews.
 E.g. in \cite{2} the Bethe--Salpeter equations have been exploited to formulate the instant form (IF) of the
 hydrogen atom Hamiltonian and the Lorentz contracted (LC) form of the wave function was found with the accuracy
$O(\alpha)$. Similar conclusions of the approximate LC wave functions  have been obtained in \cite{3,4,5,6,7}. In general one should understand e.g. what kind of approximation is the instant interaction form for the QED and QCD, and how to construct the basic Hamiltonian equation  in the IF in the c.m. system.
  We start with the instant form dynamics and the resulting hadron masses and wave functions in the c.m. system
  and then discuss how these wave functions are modified by the Lorentz boosts. Actually the hadron wave functions
  and masses are to be found in the formalisms based on the Poincare invariant forms. The most relevant and strictly relativistic IF formalism was developed on the basis
  of the relativistic path integral methods of Feynman \cite{11} and Schwinger \cite{12} and the Fock functionals \cite{13} and was called the Fock--Feynman--Schwinger method \cite{14,15,16,17}. Its application has allowed to build the standard Regge trajectories for hadron masses \cite{18,19,20} which now cover the majority of known meson states \cite{21,22,23}. As a result one has a reliable IF Hamiltonian and the rest frame wave functions which successfully correspond to the experimentally known hadrons.
  The resulting Green's functions are reproducing the fundamental results of Schwinger \cite{24}.
  To test the accuracy of this instantaneous formalism one can use the QED instantaneous Hamiltonian to calculate
  the spectrum of relativistic hydrogen-type atoms and positronium, which was performed in \cite{25}, where it was shown that the accuracy of the positronium spectrum was around $0.1 \alpha^4$. This formalism is sufficient to consider matrix elements of hadron wave functions in the same Lorentz frame, e.g. in the rest frame.

  The analysis of the same IF Hamiltonian in the moving frame was performed in \cite{26} in the case of QED and QCD interactions,where it was shown how the IF Hamiltonian transforms with the c.m. velocity, and was found that the hadron and hydrogen wave functions are Lorentz contracted. A further analysis of this QCD Lorentz contraction for the parton distributions was done in \cite{27,28}. And here we are coming to the basic point of this paper-how to calculate the matrix elements with wave functions moving with different velocities. Therefore the main emphasis of the present paper is the calculation
 of the nonlocal vertices containing hadron wave functions and its dependence on the (possibly large) momentum transfer $Q$  written in the instant form, where hadrons may have their Lorentz contracted forms. Assuming an exact
 Lorentz contracted form for the wave functions participating in the same matrix element with different velocities
 it was found that meson form factors well reproduce experimental data \cite{29},and the same is true for the baryon form factors at least for moderate Q values \cite{30}. Finally the same LC effect helps to understand the strong
  $\rho$ decays \cite{31}. It is a purpose of this paper to understand better the effects of LC in different physical systems  in the whole range of $Q$ values and for this purpose we consider the
  case of the hadron-hadron scattering with the one hadron exchange. This mechanism was considered
  recently in the case of the high energy $pp$ an $p\bar p$ scattering \cite{31*} and it was shown that treating
  hadron exchange vertices as proton form factors one can explain the observed in many experiments
 (see \cite{31*} and refs therein) the picture of two exponential slopes of the differential cross section. One may wonder whether the LC effect strongly modifying the hadron form factors at large $Q$
 is also effective in the case of hadron-hadron scattering and how to reconcile this fact with the
 strong exponential decays of the corresponding cross sections. We show below that the situation with
the h-h scattering is more complicated and the LC mechanism starts to work only near the backward direction in the hadron-hadron scattering, establishing the sharp increase of the backward differential cross section.
The plan of the paper is as follows. In section 2 is given the qualitative discussion of the perturbative vs LC mechanisms. In section 3 we discuss the relativistic transformation laws for the bound state wave functions of two and three particles, in section 4 the explicit form of the Hamiltonian and wave functions in the boosted instant form is discussed. In section 5 we discuss the LC mechanism in meson electromagnetic form factors and meson decay amplitudes. Section 6 is devoted to the high $Q$  asymptotics of the scattering amplitudes with the particle exchange mechanism. The final section contains a short summary of results and an outcome.

  \section{The qualitative discussion of the perturbative vs Lorentz contraction mechanism}

   We start with the discussion of the
  hadron form factor, considered in \cite{29,30}, which we write in a simplified form without spin or
  flavor variables
\be
F(Q)= \int d^3 q \phi_{\veP}(\veq) \phi_{\veP+\veQ}(\veq +\veQ),
\label{eq.1} \ee

Here $\phi_{\veP} $ denotes the wave function moving with the momentum $\veP$ before the collision with the photon
of the momentum $\veQ$ and the wave function after collision is moving with the momentum $\veP + \veQ$, so that the integral contains the overlap of two differently Lorentz contracted wave functions, and in addition taken at different points in the momentum space: $\veq$ and $\veq +\veQ$. This large difference of two points for large $Q$
calls for the use of the perturbation theory for the computation of the wave function at large $\veQ$, as it was done
in \cite{32,33,34,35}. In this case the wave function can be taken as a combination of $q\bar q$ or $qqq$ quarks
with one or two gluon exchanges and no question arises about the form of the wave function at all. However this
assumption of the necessary use of the perturbation theory with the free wave function components  is not obligatory, and
as it is discussed in \cite{35,36} and compared to the proton data in \cite{36*} can contradict experimental data. Indeed, as was stressed in \cite{36}, the perturbative
contributions e.g. in the proton form factor with the one- and two-gluon exchange should have a relative magnitudes
 compared with the no exchange part as $ x=\frac{\alpha_s}{\pi}$ and $x^2$ respectively, i.e. of the order of $0.1;0.01$ which  not well agrees with data and one is to look for a nonperturbative (or ``soft'' \cite{37,38,39,40} ) mechanism of high momentum transition. In this respect it is interesting what happens with the momentum dependent wave function when it experiences a Lorentz boost. As it was discussed in \cite{2,3,4,5,6,7,8,9,10,24,25,26,27,28,29,30,31} and below in this paper,the x-space
 and p-space (pseudo)scalar $q\bar q$ wave functions moving with the velocity $v$ have the form

\be
\psi_v(\vex)= \psi_0\left(\vex_\bot, \frac{ x_{\|}}{\sqrt{1-v^2}}\right),~
\phi_v(\vep)= C_0 \phi_0(\vep_\bot, p_{\|} \sqrt{1-v^2}),~ C_0= \sqrt{1-v^2}.
\label{eq.2} \ee

Now it is easy to see that the Lorentz contracted p-space wave function entering in the form factor under the integral sign will have in its argument the high momentum $Q$ multiplied by $\sqrt{1-v^2}=\sqrt{\frac{M}{M^2 +
(\veP + \veQ)^2}}$ and hence at large $Q$, $Q \gg M$, this argument is tending to the constant limit.
E.g. in the Breit frame where $\veP= -\frac{\veQ}{2}$ this upper limit is $2M$, and the wave function is not in the asymptotic region,may be large and make this mechanism preferable over the perturbative one.
This situation suggests a completely different picture of high momentum transfer for the form factors and for any
other process where wave functions in the matrix elements enter at different velocities: strong decays with a high
energy release, scattering at high $ Q$ with one or two particle exchanges, high $Q$ particle creation etc.
 In what follows in the paper we shall discuss these processes and compare our predictions with known data.

 \section{Lorentz transformations of wave functions and matrix elements}

 As it is known, there are 3 dynamical forms of relativistic interacting systems \cite{1} 1/ light front (LF), 2/ instantaneuos form (IF), 3/point-like form. One can argue that these forms are a matter of convenience and indeed in
 \cite{41} the detailed analysis has shown a close correspondence of LF and IF. In what follows we shall exploit
 only IF as it is most important for the study of bound states in QCD and QED. Indeed, in the case of QED the use of the
Bethe--Salpeter equation \cite{42} with  independent time for each particle leads to wrong additional solutions and one is
to solve instead the Salpeter equation \cite{42} based on the IF. Moreover in the case of QCD the basic interaction is confinement \cite{43} which has a global (not point-to-point as the one- gluon exchange) character and it is natural to treat systems coupled by
confinement in the IF. Below we shall describe the two and three body systems with an arbitrary total momentum $\veP$
in the IF, using the standard formalism developed in \cite{8,9,10}.
We start with the relativistic wave function of two point-like objects with total mass $M$, total momentum $\veP$ and angular momentum $\veJ$
 $$
 \Psi(x_1,x_2)= \left\{0|\phi_1(x_1) \phi_2(x_2)|M,\veJ,\veP\right\}= $$\be =\exp\left(-i P\frac{x_1+x_2}{2}\right) S_1(L_P) S_2(L_P)
 \left\{0|\phi_1(0,\ver/2) \phi_2(0,-\ver/2|M,\veJ,0\right\}.
 \label{eq.3} \ee
 Here $L_P$ is a pure Lorentz boost, and $S(L)$ is a unitary Poincare transformation
 \be
 S_{\alpha\beta}(L_P) \phi_\beta(\vex)= U^{-1}(L_p) \phi_\alpha (L_P \vex) U(L_P).
 \label{eq.4} \ee
 Specifically in the spinor case one has for $S(L)$
 \be
 S(L_P)= \exp\left(-i\frac{\omega_{\mu\nu} S^{\mu\nu}}{2}\right),~ S^{0i}=\frac{i}{4}[\gamma_0,\gamma_i],
 S^{ij}= \frac{1}{2} e_{ijk}\Sigma^k.
 \label{eq.5} \ee
 Introducing now the two-point wave function in the simplest scalar case one obtains the IF  wave function in the
 coordinate space with total momentum $\veP$
 \be
 \Psi_P(x_1,x_2)= \exp(i\veP \veR) \Psi_0\left(\ver_\bot,\frac{r_{\|}}{C_0}\right).
 \label{eq.6} \ee
 and in the momentum space

 \be
 \Psi_P(q)= C_0 \Psi_0(\veq_\bot, q_{\|} C_0).
 \label{eq.7} \ee
 Here $C_0= \sqrt{1-v^2}= \sqrt{\frac{M^2}{M^2 + \veP^2}}$.
 These last equations illustrate the main point of this paper- the Lorentz contraction at large $\veP$ stabilizes
 the the internal parallel momentum dependence  of the wave function, i.e. makes it insensitive to the large momentum
 transfers. As a result these large $Q$ effects are defined by the external factors $C_0$ and do not depend on internal dynamics.

 \section{The c.m. Hamiltonian and wave functions in the instant form of QCD and QED dynamics}

 We start with the one scalar particle in the field $A_\mu(x)$ and write the Green's function as a path integral in the $4d$ Euclidean space-time with proper time $s$

 \be
 g(x,y)= \left(\frac{1}{m^2 - D_\mu^2}\right) = \int^\infty_0 ds (D^4z)_{xy} \exp( -K) \Phi(x,y),
 \label{eq.8} \ee
 where $D_\mu= \partial_\mu - ie A_\mu, K= m^2 s + \frac{1}{4} \int^\infty_0 d\tau \left(\frac{dz_\mu}{d\tau}\right)^2,$ $
 \Phi(x,y)= \exp (ie\int^x_y A_\mu dz_\mu$,
 Here $(D^4 z)_{xy}$ is the path integral element over all random paths in $4d$.
 The important quantity $\omega$ as a virtual quark energy can be  introduced instead of the proper time $s$ when we connect it to the actual
 (Euclidean) time $t=x_4$  as follows $ds= \frac{dt}{2\omega}$ or for the whole interval $s=\frac{T=x_4- y_4}{2\omega}$, and the Green's function acquires the form
 \be
 g(x,y)= T \int^\infty_0 \frac{d\omega}{2\omega^2} D^3z \exp{-K(\omega)} \Phi(x,y),
 \label{eq.9} \ee
 where $ K(\omega)= \int^T_0 dt_E (\frac{\omega^2 +m^2}{2\omega} + \frac{\omega}{2} (\frac{d\vez}{dt_E})^2)$
 One can associate the integral in (\ref{eq.4}) with the one particle Hamiltonian as follows
 \be
 \int(D^3z)_{\vex\vey} \exp(-K(\omega))= \lan\vex|\exp(-H(\omega) T)|\vey\ran,
 \label{eq.10}\ee
 where the Hamiltonian is
 \be
 H(\omega)= \frac{\vep^2 + m^2 +\omega^2}{2\omega} + V(\vex).
 \label{eq.11}\ee
 Here $V(\vex)$ is generated by $\Phi(x,y)$, as will be shown below in the realistic case of $q\bar q$ bound state.

As a result we obtain the following expression of the one-particle Green's function via Hamiltonian
\be
g(x,y)= \sqrt{\frac{T}{8\pi}}\int^\infty_0 \frac{d\omega}{\omega \sqrt{\omega}} \lan\vex|\exp(-H(\omega) T)|\vey\ran.
\label{eq.12}\ee
From (\ref{eq.12}) one can find the value of $\omega $ using the stationary point method, e.g. in the simple case
of $V=0$ in (\ref{eq.11}) one obtains $ \omega^{(0)}= \sqrt{\vep^2 + m^2}$, i.e. $\omega^{(0)}$ is an effective quark energy.

In a similar way the $q\bar q$ Green's function can be written via the corresponding Hamiltonian
\be
G(x,y)= \frac{T}{8\pi} \int^\infty_0 \frac{d\omega_1}{\omega_1^{3/2}}\int^\infty_0 \frac{d\omega_2}{\omega_2^{3/2}} Y_\Gamma
\lan\vex|\exp(-H(\omega_1,\omega_2,\vep_1,\vep_2)) T|\vey\ran,
\label{eq.13} \ee
where $H$ is the $q\bar q$ Hamiltonian with the virtual quark energies $\omega_i, i= 1,2$
\be
H= \sum_i{\frac{\vep_i^2 +m_i^2 +\omega_i^2}{2\omega_i}} + V_0(r) + V_{SE},
\label{eq.14 } \ee
where $V_0(r)$ contains confinement and gluon exchange potentials and $V_{SE}$ is the self-energy correction \cite{44}
The resulting $q \bar q$ wave functions  are found from the equation $H \psi(\ver)= M(\omega) \psi(\ver)$ and the final
mass and wave function are found from the integral over $d\omega_1 d\omega_2$ yielding the stationary points
$\omega_i^{(0)}$. As it was shown in the QED case (hydrogen atom and positronium) this Hamiltonian formalism yields
the accuracy of eigenvalues of the order of $O(\alpha^4)$. The resulting spectra of $q\bar q$ hadrons,
\cite{21},
namely light \cite{22}, heavy \cite{23}, and heavy-light mesons \cite{23*} are in good agreement with experimental data. Note, that this formalism does not contain fitting constants except for the string tension $\sigma$ and the strong coupling $\alpha$ (e.g. $V_{SE}$ is computed via $\sigma)$. Now we are in a position to discuss the properties of the solutions $\psi(r),
\bar\psi(p)$ in the whole domain of the $p$ values, which are important for the form factors and the decay widths.
Formally one can use the rigorous expansion in the Laguerre polynomial  series with oscillator exponentials
$\psi(p)= c_n R_{nl}(\beta,p)$,where $R_{nl}$ is
\be
R_{nl}= c_{nl} \left(\frac{p}{\beta}\right)^l \exp\left(-\frac{p^2}{2\beta^2}\right)
L^{l +\frac{1}{2}}_{n-1}\left(\frac{p^2}{\beta^2}\right).
\label{eq.15} \ee
This SHO approximation was checked in \cite{45} for the heavy, heavy-light and light mesons and the resulting
accuracy of the leading term for $1S,2S$ states was around $5\%$ for heavy and heavy-light mesons. and somewhat worse for light mesons.

\section{ Boosted hadron wave functions in form factors and strong decays}

We consider below the $q\bar q$ and $3q$ IF wave functions corresponding to the meson and baryon bound states
\be
\Psi^{(2)}_{\veP} (x_1,x_2)= \lan 0|\bar \psi(\vex_1,x_0) \Gamma \psi(\vex_2,x_0) |J,\veP\ran,
\label{eq.16} \ee
\be
\Psi^{(3)}_{\veP} (x_1,x_2,x_3)= \lan 0| C_{\alpha,\beta,\gamma} \psi_\alpha(\vex_1,x_0) \psi_\beta(\vex_2,x_0)
\psi_\gamma(\vex_3,x_0)|J, \veP\ran.
\label{eq.17}\ee

We are interested in the $\veP$ dependence of the wave functions of (\ref{eq.16}), (\ref{eq.17}), which can be written in the first case as \cite{8,9,10}
\be
\Psi^{(2)}_{\veP }(x_1,x_2)= S_1(L_P) S_2(L_P) G(L_P) \Psi^{(2)}_0 ( \ver).
\label{eq.18} \ee
For the spin $\frac{1}{2}$ wave functions the transformation law is
\be
S(L_P)= \exp\left( -\frac{i}{2} \omega_{\mu\nu} S^{\mu\nu}\right),~ S^{0i}= \frac{i}{4} [\gamma_0,\gamma_i],~
S_{ij}= \frac{1}{2} e_{ijk} \Sigma^k.
\label{eq.19} \ee
and for the pure boost without rotation one has $S_0(L_P)= \sqrt{\frac{E(\veP) + M}{2 M}}$.
In the simplest case of the (pseudo)scalar bound state wave function boosted with the momentum $\veP$  acquires
the form
\be
\Psi_{\veP} (x_1,x_2)= \exp{i(\veP \veR)} \Psi_0 (\ver_ \bot, \frac{r_{\|}}{C_0} ), C_0= \sqrt{1-v^2}=
\sqrt{\frac{M^2}{M^2 + \veP^2}},
\label{eq.20} \ee
which implies the Lorentz-contracted form of the bound state. Here $\veR,\ver$ are the c.m. and relative coordinates, which are defined via the energies of the quark and antiquark. In the strong interacting $q\bar q$ system this definition includes the interaction, $ \ver= \vex_1 - \vex_2, \veR= \frac{ \omega_1 \vex_1 + \omega_2 \vex_2}
{\omega_1 + \omega_2}$. Similarly in the momentum space one obtains in this (pseudo)scalar case
\be
\Psi_{\veP} (\veq)= C_0 \Psi_0 ( \veq_{\bot}, q_{\|} C_0).
\label{eq.21} \ee
We turn now to the $3q$ wave function and consider the simplest case when the $3q$ bound state has the total spin
$\frac{1}{2}$ while the spin-dependent interaction can be neglected at the first stage,so that the total wave function undergoes the transformation
\be
\Psi^{(3)}_{\veP }(x_1,x_2,x_3)= S_0(L_P) \exp{i\veP\veR} \Psi\left(\vexi_{\bot},\frac{\xi_{\|}}{C_0},\veta_{\bot},\frac{\veta_{\|}}{C_0}\right).
\label{eq.22}\ee

Here $\vexi,\veta$ are two relative coordinates of $3q$ system, and (\ref{eq.9}) tells that all relative coordinates are Lorentz contracted in the same way as in the $q\bar q$ case. Based on that in the paper \cite{7}
the authors have predicted the squared $C_0$ behavior of the $3q$ momentum wave function for baryons (and
respectively $N-1$ power for the $N$-body wave function). However we shall now show that this situation depends on
the  type of measurement done in the $N$-body system. Namely in the form factor one is measuring the one-particle
distribution (density) in the system and as will be seen in this case for any number of constituents one obtains the
same factor $C_0^1$, while for the complicated density correlations the power of $C_0$ may be larger.
To illustrate this situation we shall derive the Lorentz contraction of the wave function in the way it was
done in \cite{29,30} for the $q\bar q$ wave function, but now we extend this derivation to the case of arbitrary constituents. One can write the for the standard one-point density the relation known from the relativity theory \cite{46}
\be
\rho(\vex,t)dV = {\rm invariant},\label{eq.23}
\ee
where $\rho(\vex,t)$ is the density, associated with the wave function $\psi_n(\vex,\vey,..,t)$,
\be
\rho_n(\vex,t) = \int\left(d\Gamma \frac{1}{2i} \left(\psi_n \frac{\partial \psi_n^+}{\partial t}  - \psi_n^+ \frac{\partial \psi_n}{\partial t}\right)\right)
= E_n\int(d\Gamma |\psi_n(\vex,t)|^2), ~~\label{eq.24}
\ee

and $dV=d\vex_{\bot} dx_{\|}$,while $d\Gamma$ implies the integration over all other coordinates $\vey,\vez,..$ on both sides in the Lorentz invariant way,e.g. in the moving frame $d\Gamma$ means $d\vey_{\bot}dy_{\|}\frac{1}{C_0}$ and so on for other extra coordinates. One can use the standard transformations,
\be
L_{\rm P}dx_{\|} \rightarrow dx_{\|} \sqrt{1 - \vev^2}, ~~ L_{\rm P} E_n \rightarrow \frac{E_n}{\sqrt{1-\vev^2}},
\label{eq.25}
\ee
to insure the invariance of  (\ref{eq.10}). In its turn the invariance law implies that in the wave function $\psi(\vex,t)=\exp(-iE_nt)\varphi_n(\vex)$ the function $\varphi_n(\vex)$ is deformed in the moving system,
\be
L_{\rm P}\varphi_n(\vex_\bot, x_{\|},\vey_\bot, y_{\|},..) = \varphi_n\left(\vex_\bot, \frac{x_{\|}}{C_0},\vey_\bot,\frac{y_{\|}}{C_0}\right)
\label{eq.26} \ee
One can see that the integration of all extra coordinates is performed in the Lorentz invariant way.
Going now to the momentum dependent $3q$ wave functions one obtains as in (\ref{eq.8}) one power of $C_0$ for the Lorentz deformed basic momentum $k_{\|}$ and Lorentz invariant integration over all other momenta, so that the general single point form factor of the multiparticle system has the form (in the Breit system $\veP_1 +\veQ= \veP_2,
\veP_1= -\frac{\veQ}{2}$
$$
F^{(1)}(Q)= (C_0)^2 \int(d\vek_{\bot} dk_{\|} d\veq_{\bot} dq_{\|} C_0 d\veq'_{\bot} dq'_{\|} C_0 ...
\Psi_0(\vek_{\bot},k_{\|}C_0,\veq_{\bot},q_{\|}C_0),...)$$
\be
V \Psi_0((\vek_{\bot},(k_{\|} + a_1 Q)C_0, \veq_{\bot},
(q_{\|} + a_2 Q)C_0,...,
\label{eq.27}\ee
Here $C_0= {\frac{2M}{\sqrt(4M^2 + Q^2)}}$, and $a_1,a_2$ define how the relative momenta $k,q$ are connected with
the momentum of the selected particle. One can now introduce the variables $\bar k= k_{\|} C_0, \bar q= q_{\|} C_0, etc$. As a result the single point form factor can be written as
\be
F^{(1)}(Q)= C_0 \int( d\vek_{\bot} d\veq_{\bot}... d\bar k d\bar q ... \Psi_0(\vek_{\bot},\bar k,\veq_{\bot},\bar q,...)V \Psi_0(\vek_{\bot},\bar k+ Q C_0,\veq_{\bot},\bar q+ Q C_0,...)).
\label{eq.28}\ee

As a result one can see two main features, one is basic for all hadron matrix element: the appearance of the argument $Q C_0= \frac{Q 2M}{\sqrt{Q^2 + 4M^2}}$ which is finite ($2M$) in the large $Q$ limit  is basically different
from the perturbative results \cite{32,33,34}. The second property of (\ref{eq.15}) is the factor $C_0$ in front of the
integral,which ensures the $Q^{-1}$ behavior of form factors for any number of constituents. Note however that the power of $C_0$ depends on the phenomenon under study and
can change e.g. in the case of inelastic form factors, cross sections etc., which will be discussed below in the paper.
To make the role of the LC in the form factors more visible one can write the form factor of the PS mesons
in an explicit form

\be
F(Q^2)= C_0 \int \frac{d^2\veq_\bot d\kappa}{(2\pi)^3}\phi\left(\veq_\bot,\kappa)\phi(\veq_\bot,\kappa +
\frac{Q m_h}{2\sqrt{m_h^2 + Q^2/4}}\right).
\ee \label{eq.28'}
As it was shown in \cite{29} in the simplest case when one can consider meson wave functions of the oscillator type one obtains a one parameter equation for the meson form factor
\be
F(Q^2)= \frac{m_h}{\sqrt{m_h^2 + \frac{Q^2}{4}}} \exp\left(-\frac{Q^2 m_h^2}{16 k_0^2 (Q^2/4 + m_h^2)}\right).
\label{eq.28"}\ee
Here $k_0$ is the oscillator radius of the momentum wave function. In \cite{29} examples of the $K$ and $\pi$ form factors are given in comparison with the lattice data showing a reasonable agreement. Below we shall
do the same for the $\eta_c$ form factor using \ref{28"} with the parameter $k_0= 0.7$ GeV in comparison with the lattice data from \cite{46'}. The results of this comparison are given below in the Table 1.

\begin{table} [!htb]
\caption{Scalar form factor of $\eta_c$ meson in comparison with the lattice data \cite{46'}}
\begin{center}
\label{tab.01}
\begin{tabular}{|l|c|c|c|c|c|}
\hline
$Q^2$(in GeV$^2$) & 1 & 5 & 10 & 15 & 20\\
$Q^2 F_{\eta_c}$ (th) &0.87 &2.67 &3.69 &3.26 &3.3\\
$Q^2 F_{\eta_c}$(lat) &0.87 &2.7 &3.6 &3.8 & 4\\
\hline
\end{tabular}
\end{center}
\end{table}
One can see a reasonable agreement of the theory and lattice data with the use of the only parameter-
the momentum radius of the wave function $k_0= 0.7 $ GeV.

Even more interesting is the situation in the proton and neutron form factors studied in \cite{31}.
Here again one can use the oscillator form with the only parameter of the nucleon momentum radius $\lambda$ for all 4 form factors of the proton and neutron and the additional parameter $A$ characterizing the admixture of the P waves in the total nucleon wave functions. It is shown in \cite{31} that all 4 form factors describe well the data below 12 GeV, which is especially important for the electric neutron
form factor which turns to zero in absence of the P-waves when $A=0$. Below we show the additional comparison of the theory and experimental data from \cite{46"} in a wider $Q^2$ interval $3-16$ GeV$^2$.
The resulting equation for the proton magnetic form factor contains the same two parameters $A,\lambda$
\be
\frac{G^p_M}{\mu_p}= \bar f(Q)\left (1 -\frac{1}{3}\sqrt{\frac{2}{3}} \frac{A}{6\lambda^2} g(Q^2)\right), \label{eq.28'''} \ee
Where $\bar f(Q)= C_0 \exp\left(-\frac{C_0^2 Q^2}{6\lambda^2}\right), g(Q^2)= Q^2 C_0^2$.
One can see here the effect of the LC-everywhere the factors $Q^2$ are multiplied by the LC factors $C_0^2$.
The resulting form factor theoretical values are compared below with the experimental data of \cite{46"}.
\begin{table}[!htb]
\caption{Proton magnetic form factor $\frac{Q^4 G^p_M}{\mu_p}= P(Q)$ from \ref{eq.28'''} in comparison with the
experimental data of \cite{46"}}
\begin{center}
\label{tab.02}
\begin{tabular}{|l|c|c|c|c|c|}
\hline
$Q^2$ (in GeV$^2)$ &2.862 &5.02 & 9.629 &11.99 & 0.378\\
$P(Q)$(exp)      &0.331 &0.390 & 0.390 & 0.392 & 0.378\\
$P(Q)$(theor)    &0.342 &0.377 & 0.443 & 0.46  & 0.54\\
\hline
\end{tabular}
\end{center}
\end{table}
As one can see in this table the agreement with experiment on the level of $0.05-0.1$ except at largest
$Q^2$, possibly requiring additional details of the proton wave function.

Another example of the effect produced by the LC phenomenon is the strong decay of a hadron with fast d
ecay products, discussed in \cite{30}. Here the decay width can be written via the decay matrix element $J(E)$ of a static decaying hadron $\Psi^{0}$ into moving hadrons $\psi^{v}_{1,2}$ as follows
$$
\Gamma(E)={\rm const }~p(E)^{2L+ 1} |J(E)|^2, J(E)= {\rm const} \int d^3 q \Psi^{(0)}_1(E,\veq)\psi^{v}_2(q)\psi^{v}_3(q)=$$$$=
{\rm const}~ C_0^2 \int d\veq_{\bot} dq_{\|} \Psi^{(0)}(\veq) \psi^{0}_2(\veq_{\bot},q_{\|}C_0) \psi^{0}_3(\veq_{\bot},q_{\|}C_0)=$$\be= {\rm const}~ C_0 \int d\veq_{\bot} d\kappa \Psi^{0} \psi^{0}_2(\veq_{\bot}\kappa)
\psi^{0}_3(\veq_{\bot},\kappa),
\label{eq.29} \ee
and as a result one obtains for the decay with high energy release
\be
  \Gamma(E)= C_0^2 \Gamma(2m)= \frac{4m^2 \Gamma(2m)}{s}.
 \label{30} \ee
 This $s=E^2$ reduction  was known experimentally for decades \cite{47} and remarkably the LC can explain it.

\section{Boosted hadron wave functions in hadron-hadron scattering}

In the previous sections we have considered reactions where  the boosted hadron wave functions have been
functions of one boost direction, as form factors in the Breit system or decay products of the fixed resonance at rest. Here we shall consider the hadron-hadron scattering in the c.m. system with exchange
of a hadron. We shall assume that the interaction is due the one-hadron-exchange process, and the basic
point is that each exchange vertex is proportional to the overlap integral of the same type as in the
form factor but now in the general frame. In the case of the $p-p$ scattering in the c.m. system one can
write initial and final momenta as $\vep,-\vep$ and $\vep', -\vep'$ and $\veQ= \vep- \vep'$.
Following \cite{31*} the total scattering amplitude can be written as
\be
A = G_1(s,t) + G_2(s,t) = g_1(s,t) F_1(t) + g_2(s,t) F_2(t),~~ t= -\veQ^2,
\label{eq.31}
\ee

where $g_i(s,t)$ may contain one-boson exchange (OBE) pole, or a branch point, and multiple s-dependent corrections, while $F_i(t)$
is a product of two form factor vertices, containing the momentum transfer to the proton in the vertices (1,3) and (2,4), $F_i(t)= f^i_{13}(Q) f^i_{24}(Q)$, where

\be
 f_{13}(Q): P_1(\vep) \to P_3(\vep') + b(\veQ);~~ f_{24}(Q): P_2(-\vep) + b(\veQ) \to P_4(-\vep').
\label{eq.32}
\ee

At this point one should stress the difference between the electromagnetic form factor of the proton
and the form factors $f^i_{13}$.
Both $f^i_{13}, f^i_{24}$ for $i=1$ correspond to the one-quark interaction with the exchanged hadron and as shown in \cite{31*}are equal to the proton form factor $f_(Q)$  which can be written via the proton wave functions as

\newcommand{\velambda}{\mbox{\boldmath${\rm \lambda}$}}
\be
f_i(Q) = \sum_k \kappa_k \int d^3 \vexi d^3 \veta \psi(\vexi,\veta) \exp(i\veQ \velambda^{i}_k) \psi(\vexi,\veta).
\label{eq.33}
\ee

In the case $i=2$ the exchanged hadron contacts the c.m. of two quarks in both vertices and the resulting form factors
are again expressed via electromagnetic ones with additional coefficient in the exponent, see \cite{31*}.
As a result two vertices $f_1, f_2$ in the one-hadron-exchange process are
\newcommand{\vea}{\mbox{\boldmath${\rm a}$}}
\be
f_1(Q,p,p')= \int d^3\xi d^3\eta \psi_{\vep}(\vexi,\veta) \exp(i\veQ \vea)\psi_{\vep'}(\vexi,\veta).
\label{34}\ee
Here $\vea= \nu_1 \vexi + \nu_2 \veta$, and $\nu_i$ are defined by the exchange process discussed above-whether the exchanged hadron couples to one or two quarks in the baryon-see the explicit derivation in \cite{31*}.
Following this derivation for $f_2(Q,p,p')$ one obtains the same expression as in (\ref{eq.31}) but with
the replacement $\nu_1 , \nu_2$ into another set of values.
It is important that each vertex function $f_i(Q,p,p')$ can be defined as the Lorentz invariant function for arbitrary values of $Q,p,p'$ as it is done in the Appendix. In the typical case of the
nearly forward scattering one can consider the case when $Q$ is much smaller than both $p$ and $p'$
and can consider the system moving with the momentum $\veP= \frac{\vep + \vep'}{2}$ which is different from the well-known Breit system used for e.m. form factors. Namely, in the Breit system the only direction for
initial and final momenta is the vector $\veQ$ and therefore the contraction occurs in this direction,
yielding the final $Q$ dependence as $Q C_0(Q)$ which is const at large $Q$.
Introducing now the momentum space wave functions $\psi(\vexi,\veta)= \int \frac{d^3 q d^3 k}{(2\pi)^6} \exp{i(\veq \vexi + \vek \veta)} \phi(\veq,\vek)$, one can approximately neglect at $Q<<P$ the contraction along $\veQ$ as compared to the strong contraction along $\veP$. As a result $\veQ$ adds to the transverse momenta $\veq, \vek$ and avoids contraction coefficient $C_0(P)$. One obtains
\be
f_1(Q)= C_0^2 \int \frac{d^3 q d^3 k}{(2\pi)^6} \phi(q_{\|} C_0,\veq_\bot,k_{\|}C_0,\vek_\bot)
\phi((q_{\|} C_0,\veq_\bot + \veQ, k_{\|}C_0, \vek_\bot + \veQ).
\label{eq.35}\ee
In (\ref{eq.35}) parallel and transverse directions are defined with respect to the vector $\veP$.
Here $C_0$ is equal to $\sqrt{\frac{4M^2}{4M^2 + P^2}}$. Assuming as in \cite{31*} the Gaussian form
of the wave functions $\phi(\veq,\vek)= {\rm const} \exp(-\frac{q^2 + k^2}{2\mu^2})$, one obtains in (\ref{eq.31})
$$
F_i(t)=F_i(Q)=f_i(Q)^2,
F_1(Q)=  \exp\left(-Q^2 \frac{2}{3\mu^2}\right)= \exp\left(-\frac{B_1 Q^2 }{2}\right),$$ \be
 F_2(Q)= \exp\left(-Q^2 \frac{1}{6\mu^2}\right)= \exp\left(-\frac{B_2 Q^2 }{2}\right).
\label{36}\ee
One can connect as in \cite{31*} the parameter  $\mu$ with the proton charge radius
$\frac{1}{\mu^2}= \frac{r_p^2}{(hc)2}= 17.18 r_p^2$, which agrees well with the pp scattering data
for $r_p^2= 0.93~ Fm^2$.
Inserting $F_1, F_2$ into (\ref{eq.31}) one can write the differential cross sections of $p-p$ scattering as in \cite{31*}
\be
\frac{d\sigma}{dt}= |g_1(s,t)|^2 \exp(B_1  t) + |g_2(s,t)|^2 \exp(B_2  t) +
2 Re(g_1 g_2^*) \exp\left(\frac{(B_1 + B_2) }{2} t\right).
\label{37}\ee
Here $B_1,B_2$ are defined as in \cite{31*}
\be
B_1= \frac{2}{3 \mu^2}= \frac{2 r_p^2}{3 (hc)^2}= 16~{\rm GeV}^{-2}, B_2= \frac{1}{4} B_1.
\label{38}\ee
Note, that this is true only in the limit $Q << P$, and therefore applies to the scattering data in the GeV or TeV region with $-t= Q^2 << s$. A more general situation is discussed in the Appendix A The opposite situation occurs in the case of the backward
scattering, which is interesting to discuss in more detail. In this case $\vep'= -\vep$, $\veQ= 2\vep$
and one has exactly the same situation as in the form factor case in the Breit system.
As a result in the case of backward $pp, p\bar p$ scattering all factors $t$ in (\ref{37}) should be replaced by $t C_0^2$ with $C_0^2= \frac{M^2}{M^2 + p^2}$ and the final expression for the backward
differential cross section acquires the form
\be
\frac{d\sigma_bw}{dt}= |g_1(s,s-4M^2)|^2 \exp(-B_1 w) + |g_2(s,s-4M^2)|^2 \exp(-B_2 w) +
2 Re(g_1 g_2^*) \exp\left(\frac{(B_1 + B_2)}{2} w\right).
\label{39} \ee
Here $w= \frac{4 p^2 M^2}{M^2 + p^2}$ which tends to $4 M^2$ at large $s$. In this way the Lorentz contraction  produces the stabilization effect for the backward scattering (modulo power decreasing
due to hadron propagators). Indeed $g_i(s,t)$ contains the hadron exchange propagator
 $\frac{1}{m_h^2 + t}$, which decreases in the case of backward scattering as $\frac{1}{s}$ and therefore one can expect the ratio of cross sections
 \be
  R(s)= \frac{\frac{d\sigma(s,t= -4M^2)}{dt}}
 {\frac{d\sigma_bw(s)}{dt}}=\frac{16 M^4}{s^2}.
 \label{40},\ee.
 Note also that this backward point in the differential cross section is a result of a steep rise.
 Indeed for nearly backward scattering at an angle $\pi -\delta$  the factor $w$ in (\ref{39})
 is replaced by $w + c p^2 (\sin(\delta))^2$.

\section{Discussion of the results and comparison to the data}

We have discussed above the dynamics of strong interacting full size hadrons in contrast to the
simplified theory with point-like objects. Three types of processes have been considered: strong
decays, form factors and hadron-hadron scattering, searching for the effects of the relativistic
Lorentz contraction of participating hadrons. It was shown that the fast LC hadrons become insensitive
to the large momentum transfer at $Q>>M$ which strongly modifies the dynamics of hadrons and in particular the large $Q$ asymptotics of form factors and decay widths. One may envisage the role of this new
phenomenon in all hadron processes with large $Q$ transferred to one or several reaction products.
The comparison of the resulting LC hadron form factors with data in the present paper (Tables 1 and 2)
and in \cite{29,31} as well as the analysis of the $rho$ decay in \cite{30} support the importance of the LC effects.
In the last topic of this paper we have considered the high energy hadron-hadron elastic scattering proceeding via one hadron exchange. In this case one must distinguish between the effects of the high
momentum of colliding hadrons and the high transferred momentum Q. For low Q values,e.g. in the
$pp$ scattering for $Q$ below $2M$, the cross section has a specific form with two slopes  which was
analyzed recently in \cite{31*} without any account of the LC wave functions.
The form of the (\ref{37}) for the forward scattering  implies  the general structure of the two-slope differential cross section which is well described by the two pomeron exchange model in \cite{46'''},while the explicit values of slope parameters are explained in \cite{31*}, i.e. it is defined by two slopes with the interference term, and the magnitude of these slopes
and the position of the connecting point were calculated and compared to the $pp$ and $p\bar p$ experimental data
in \cite{31*} finding a reasonable agreement. We have shown above that the LC effect indeed is absent in the large $p$ and small $Q$ region and is appearing only in the backward scattering region when $p$ and $Q$ are of the same order. The very idea that the backward scattering can be increased by the Lorentz contraction and the consequent suppression of the $Q^2$ dependence can be exploited in future both in theory and experiment. However the expected ratio given in (\ref{40}) can be very small for large $s$. Indeed, even for a rather low energy of colliding $pp$ or $p \bar p$
$E= \sqrt{s}= 10$ GeV the ratio is  small, $R(100 $ GeV$^2)= 0.0012$ and one needs special efforts to measure it. It i important however that also other types of mechanisms in addition to the one hadron exchange can be important in this region.
The author is grateful for useful discussions to A.M.Badalian.

\newpage

\vspace{1cm}

{\bf Appendix A}

 {\bf  The structure of the boosted wave functions in the transition matrix element}

 \setcounter{equation}{0} \def\theequation{A\arabic{equation}}

One can rewrite the matrix element $f_1(Q)$ given in (\ref{eq.35}) in the general case of arbitrary
magnitudes of $Q,p,p'$ taking into account that wave functions $\psi(\veq,\vek)$ depend on $q^2 + k^2$
as in the Gaussian form. In this way we obtain
\be
\psi_p(\veq,\vek)= C_0(p)\psi_0(\veq^2-\frac{(\veq \vep)^2}{p^2} + \frac{(\veq \vep)^2}{p^2} C_0^2(p)
+ (\veq \to \vek)).
\label{A1}\ee
\be
\psi_p'(\veq + \nu_1 \veQ,\vek + \nu_2 \veQ)= C_0(p') \psi_0((\veq + \nu_1 \veQ)^2 - \frac{((\veq + \nu_1 \veQ)\vep')^2}{p'^2} C_0^2(p') + (\veq \to \vek, \nu_1 \to \nu_2)).
\label{A2}\ee

One can see in (\ref{A2}) that at $Q$ much smaller than $p$ the first term in the argument of $\psi_0$
is dominating and the $Q$ dependence is not suppressed by the $C_0^2(p')$ factor-the Lorentz contraction effect is switched off. To make it work one must align both $\vep,\vep'$ along $\veQ$
as it happens in the backward scattering case.

\end{document}